\documentclass{nature}

\usepackage{graphicx}
\makeatletter
\let\saved@includegraphics\includegraphics
\AtBeginDocument{\let\includegraphics\saved@includegraphics}
\renewenvironment*{figure}{\@float{figure}}{\end@float}
\makeatother
\usepackage{color}

\usepackage{bm}
\usepackage[version=3]{mhchem}
\usepackage{natbib}
\usepackage{epsfig}
\usepackage{amsmath}
\usepackage{bbold}
\usepackage{float}
\usepackage{comment}

\usepackage[dvipsnames]{xcolor}

\newcommand{\RNum}[1]{\uppercase\expandafter{\romannumeral #1\relax}}

\makeatletter
\newcommand*{\rom}[1]{\expandafter\@slowromancap\romannumeral #1@}
\setcitestyle{super}

\title{Demonstration of electron-nuclear decoupling at a spin clock transition}

\author{Krishnendu Kundu,$^{1}$ Jia Chen,$^{2,3,4}$ Silas Hoffman,$^{2,3,4}$ Jonathan Marbey,$^{1,4,5}$ Dorsa Komijani,$^{1,5}$ Yan Duan,$^{6}$ Alejandro Gaita-Ari\~no,$^{6}$ John Stanton,$^{3,4,7}$ Xiao-Guang Zhang,$^{2,3,4}$ Hai-Ping Cheng,$^{2,3,4}$ Stephen Hill$^{1,4,5}$}

\begin{document}

\maketitle
\begin{affiliations}
\item National High Magnetic Field Laboratory, Florida State University, Tallahassee, FL 32310, USA
\item Department of Physics, University of Florida, Gainesville, FL 32611, USA
\item Quantum Theory Project,  University of Florida, Gainesville, FL 32611, USA
\item Center for Molecular Magnetic Quantum Materials
\item Department of Physics, Florida State University, Tallahassee, FL 32306, USA
\item Instituto de Ciencia Molecular (ICMol), Universidad de Valencia, Paterna, Spain
\item Department of Chemistry, University of Florida, Gainesville, FL 32611, USA
\end{affiliations}
\begin{abstract}
 The ability to design quantum systems that decouple from environmental noise sources is highly desirable for development of quantum technologies with optimal coherence. The chemical tunability of electronic states in magnetic molecules combined with advanced electron spin resonance techniques provides excellent opportunities to address this problem. Indeed, so-called clock transitions (CTs) have been shown to protect molecular spin qubits from magnetic noise, giving rise to significantly enhanced coherence. Here we conduct a spectroscopic and computational investigation of this physics, focusing on the role of the nuclear bath. Away from the CT, linear coupling to the nuclear degrees of freedom causes a modulation and decay of electronic coherence, as quantified via electron spin echo signals generated experimentally and \emph{in silico}. Meanwhile, the effective hyperfine interaction vanishes at the CT, resulting in electron-nuclear decoupling and an absence of quantum information leakage to the nuclear bath, providing opportunities to characterize other decoherence sources.

\end{abstract}

\section*{Main}
The synthetic tunability of molecular nanomagnets provides a versatile platform for exploring and potentially harnessing unique physical attributes that are of utility for the development of next-generation quantum information (QI) technologies.\cite{Gaita-Arino2019, Atzori2019, Godfrin2017,Bayliss2020} In particular, the electronic spin associated with a magnetic molecule may serve as the computational basis for a quantum bit, or qubit. However, as with any such system, protection from environmental noise that causes decoherence is of critical importance, representing one of the main hurdles on the path towards practical applications. In an attempt to suppress one of the more stubborn sources of decoherence arising from electron-nuclear interactions, various synthetic strategies have been employed such as nuclear spin patterning\cite{Graham2017, Jackson2019} and the use of nuclear spin free ligands.\cite{Zadrozny2015, Yu2016} However, demonstration of long phase memory (coherence) times typically still requires extreme dilution in order to minimize electron spin-spin dephasing. 

Rather than modifying the spin bath, an alternative approach involves so-called clock transitions (CTs)\cite{Bollinger1985} at which the electron spin resonance (ESR) frequency is insensitive to the local magnetic induction and, therefore, does not couple to the fluctuating magnetic environment. Such CTs occur at avoided level crossings associated with the Zeeman splitting of qubit basis states. This approach is well established in solid-state materials such as donor atoms in silicon\cite{Wolfowicz2013,Morello20} or defect states in various other host crystals. \cite{Stark18,Miao2020,Hemmer20,Probst2020,Onizhuk2021} Our interest is in molecular systems, for which enhanced coherence was demonstrated at a CT for a [Ho(W$_5$O$_{18}$)$_2$]$^{9-}$ molecule by Shiddiq et al.\cite{Shiddiq2016} Subsequently, CTs have been studied in other molecular systems\cite{Harding17,Zadrozny17,Collett19,Collett20,Kundu21} and the effects of structural distortions have been analyzed theoretically for several Ho$^{\mathrm{III}}$ and V$^{\mathrm{IV}}$ complexes.\cite{Santa2020}

\begin{figure}
\includegraphics[width=\linewidth,clip]{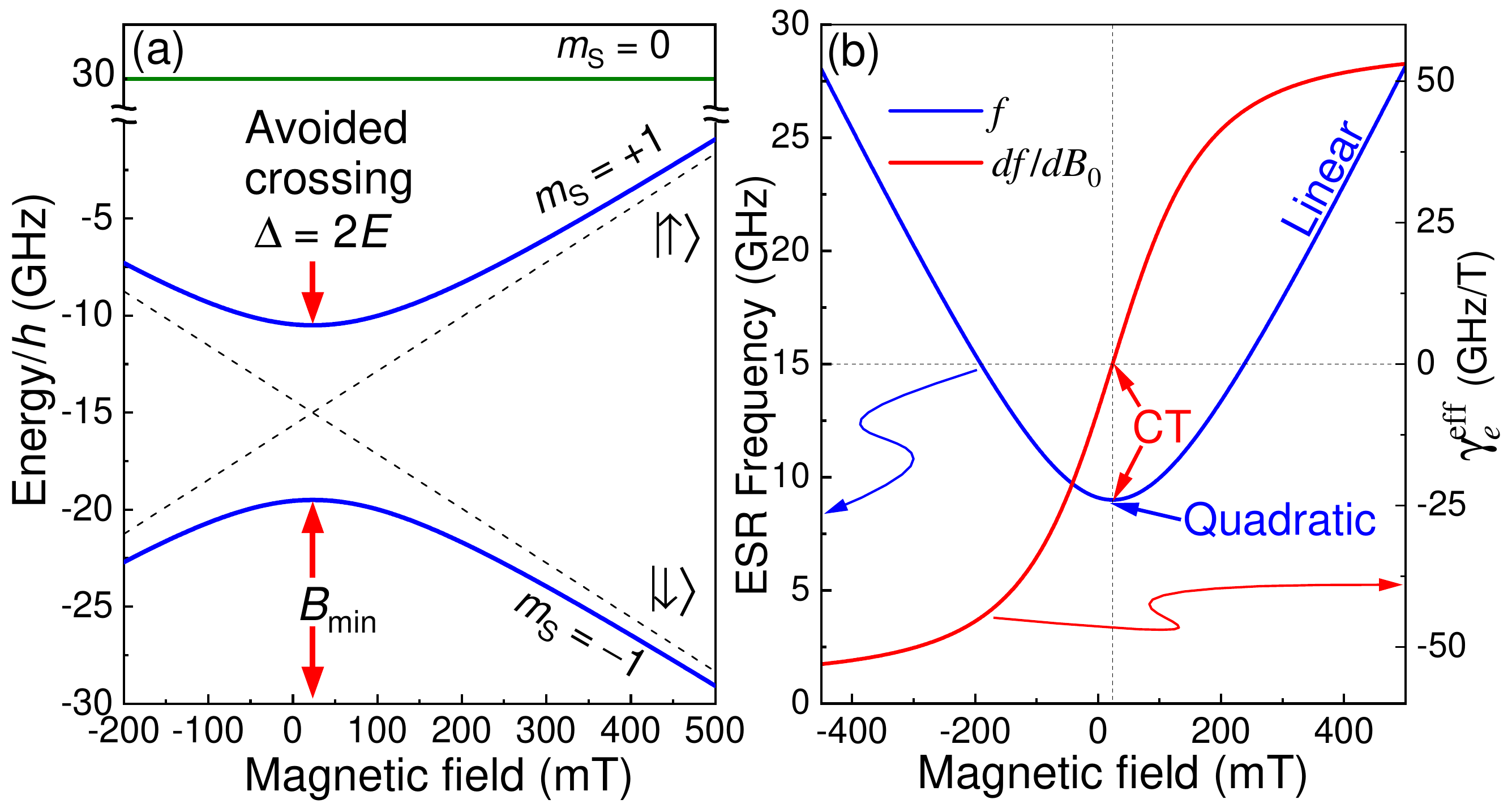}
\caption{(a) Zeeman diagram according to the Hamiltonian of Eq.~(\ref{ZH}), with the parameters given in the main text. An avoided crossing (a CT) between the two lowest lying states (blue curves - see labeling) is seen at $B_{0z} = B_{\mathrm{min}} = 23.5$~mT. (b) ESR frequency, $f$, corresponding to the (clock) transition between the $m_S=\pm 1$ states in (a), and the associated effective gyromagnetic ratio, $\gamma_e^{\textrm{eff}}=df/dB_0$. Note that the ESR frequency couples linearly to $B_0$ far from the CT and quadratically at the CT, such that $\gamma_e^{\textrm{eff}}$ crosses through zero.} \label{Zeeman}
\end{figure}

Here we directly investigate electron-nuclear coupling in the vicinity of a CT by means of pulsed electron-spin-echo (ESE) measurements and numerical modelling. Away from the CT, dipolar hyperfine coupling to the nuclear bath results in periodic modulations of the electronic coherence\textemdash the so-called ESE envelope modulation (ESEEM) effect.\cite{Schweiger01} This modulation vanishes at the CT. Theoretically, we consider a minimal model that can host a CT: an $S = 1$ spin subject to a relatively strong axial magnetic anisotropy, with an avoided Zeeman level crossing generated by a weaker transverse interaction [Fig.~\ref{Zeeman}(a)]. We treat coupling to the nuclear bath explicitly to reproduce the ESEEM effect via quantum dynamics simulations. The parameters in our simplified $S = 1$ model are chosen to mimic the low energy physics of the [Ho(W$_5$O$_{18}$)$_2$]$^{9-}$ molecule, the only system for which ESEEM has been characterized as a function of applied magnetic field, \boldmath${B}$\unboldmath$_0$, in the vicinity of a CT. The simulations compare favorably with experiment. Crucially, we demonstrate electron-nuclear decoupling at the CT. Although the experiments focus on [Ho(W$_5$O$_{18}$)$_2$]$^{9-}$, our model applies quite generally for the coupling of an electronic spin to a finite nuclear bath. The combined study provides a microscopic view of the mechanism via which an electron spin qubit couples to nearby nuclei, in essence mediating leakage of quantum information to the nuclear bath.

Pulsed ESR, which is central to most spin-based QI device implementations,\cite{Wolfowicz2016} is an extremely powerful technique enabling both sample characterization and quantum control. The simplest illustration involves the two-pulse Hahn echo sequence,\cite{Schweiger01,Hahn1950} where a coherent superposition of spin ``up" and ``down" states is first generated via a $\pi/2$ rotation on the Bloch sphere, and then the magnetization is allowed to evolve freely in the $xy$-plane; this evolution is later inverted via application of a $\pi$-pulse, ideally refocusing any dephasing that occurs due to static disorder, resulting in emission of an ESE at time $2\tau$ after the initial $\pi/2$ pulse ($\tau$ is the delay between pulses). A dynamic environment causes decoherence,\cite{Prokofiev00} which manifests as a decay of the ESE intensity upon increasing $\tau$. Meanwhile, coherent interactions with nearby quantum systems, e.g., other electrons or atomic nuclei, can give rise to a modulation of the ESE intensity.\cite{Schweiger01} In particular, ESEEM arises due to excitation of formally forbidden nuclear transitions during the pulsed ESE sequence, through hyperfine coupling to the central electron spin. Here, “central” refers to spins that have been prepared in a prescribed coherent quantum state, e.g., via application of a $\pi/2$ pulse. ESEEM may therefore be used to characterize this aspect of the environment, providing uniquely sensitive fingerprints of electron-nuclear decoherence mechanisms.

In order to gain microscopic insights into electron-nuclear coupling in the vicinity of a CT, ESEEM measurements were performed on a crystal of Na$_9$[Ho$_{0.001}$Y$_{0.999}$(W$_5$O$_{18}$)$_2$]$\cdot n$H$_2$O (hereon abbreviated HoW$_{10}$), i.e., $0.1\%$ HoW$_{10}$ doped into an isostructural non-magnetic YW$_{10}$ host. Ho$^{\mathrm{III}}$ possesses a ground state spin-orbit coupled angular momentum, $J = L+S = 8$. The pseudo-axial coordination geometry imposed on the Ho$^{\mathrm{III}}$ ion results in a crystal field (CF) interaction that lifts the degeneracy of the $2J +1$ projection ($m_J$) states, giving rise to a singlet and a series of $m_J \approx \pm i$ ($i=1$ to 8) quasi-doublets, with the $m_J = \pm 4$ ground doublet lying $\approx$~40~cm$^{-1}$ below the first excited CF states.\cite{Ghosh2012,Vonci17} A weak tetragonal CF interaction is effective in generating an avoided Zeeman level crossing between the $m_J = \pm 4$ basis states,\cite{Shiddiq2016,Liu2021} thus giving rise to a 9.18~GHz CT. The hyperfine interaction involving the $I = \frac{7}{2}$ $^{165}$Ho nuclear spin further splits the $m_J = \pm 4$ states into $(2I + 1) = 8$ pairs of $m_I$ sub-levels, resulting in eight avoided-crossings, i.e., eight CTs, four either side of zero applied field (see Fig.~1 in Ref.~[\citen{Shiddiq2016}]). For reasons explained in Ref.~[\citen{Shiddiq2016}], we focus here on the lowest field CT (at $B_{0z} = 23.6$~mT), which also gives the strongest ESEEM; note that, due to a small sample misalignment, this occurs at at $B_0 = B_{\mathrm{min}} = 25.5$~mT in the present investigation (see Methods).

\begin{figure}
\centering
\includegraphics[width=1.0\linewidth,clip]{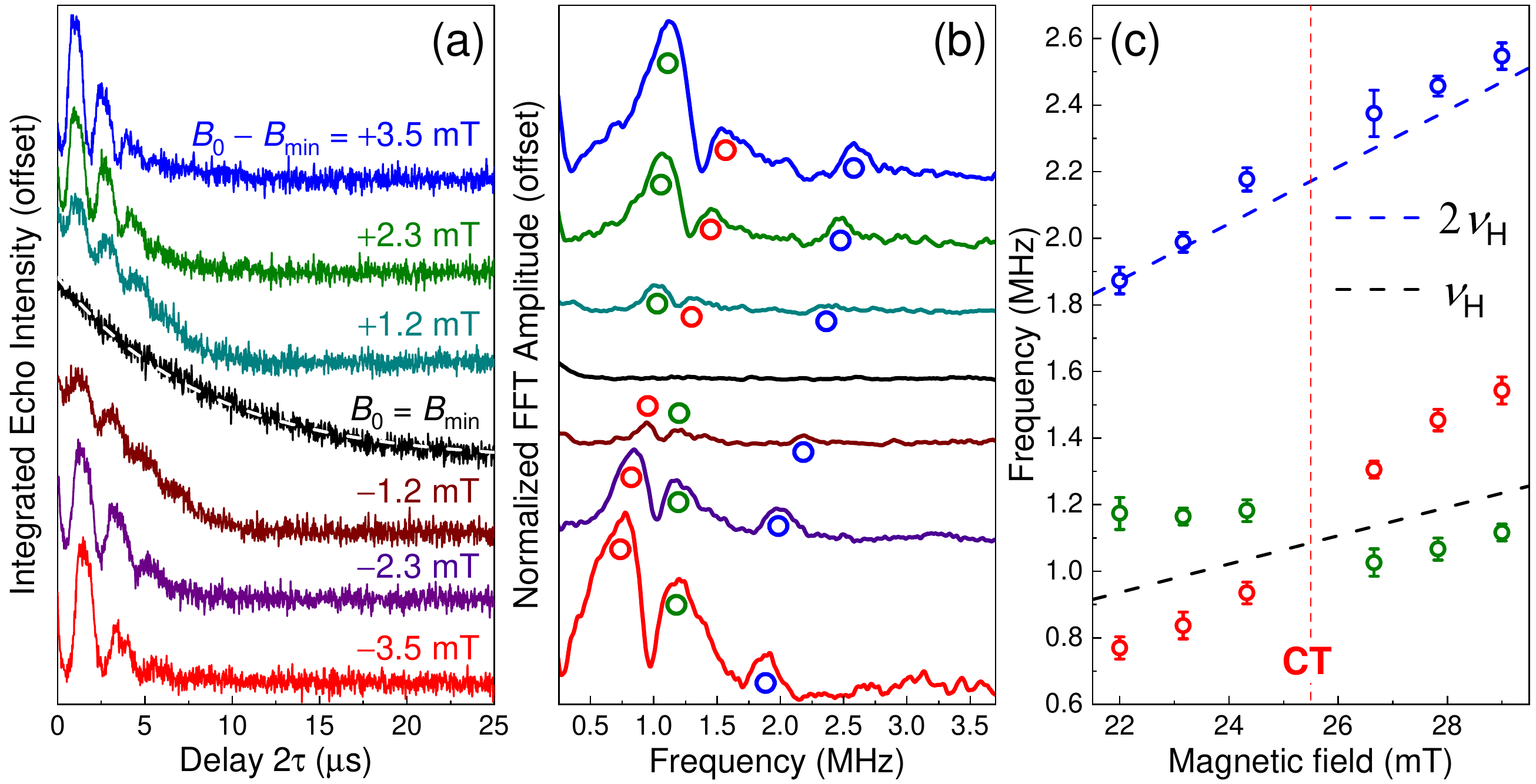}
\caption{(a) HoW$_{10}$ ESE decay curves recorded at 9.18~GHz and 5~K as a function of detuning, $B_0 - B_{\mathrm{min}}$; the white dash curve is a fit to a mono-exponential decay, from which the optimum $T_\mathrm{m}=8.43(6)~\mu$s is deduced. (b) FFTs of the decay curves in (a); prominent peaks in the ESEEM spectra are marked with red, green and blue circles. (c) Plot of ESEEM frequencies in (b) versus $B_0$; the dashed lines correspond to harmonics of the proton Larmor frequency, the open circles are colored according to the same scheme as those in (b), and the vertical red line marks the CT. } \label{exp}
\end{figure}

ESE time traces recorded at a frequency of 9.18~GHz are shown in Fig.~\ref{exp}(a) for different detuning fields ($\Delta B = B_0-B_{\mathrm{min}}$) from the CT, revealing strong temporal modulations (ESEEM) at most detunings. The first thing to note is the variation in decay time ($\equiv$ phase memory time, $T_\mathrm{m}$) and modulation depth as a function of the detuning. Most notably, there is a complete absence of ESEEM at zero detuning, i.e., at the CT. Fast Fourier transforms (FFTs) of the time traces reveal three prominent peaks, highlighted by the red, green and blue circles in Fig.~\ref{exp}(b). The associated ESEEM frequencies are plotted as a function $B_0$ in Fig.~\ref{exp}(c); superimposed on the data are the 1\textsuperscript{st} and 2\textsuperscript{nd} harmonics of the bare proton Larmor frequency, $\nu_\mathrm{H} = \gamma_\mathrm{H} B_0$, where $\gamma_\mathrm{H} = 42.577$~MHz/T is the proton gyromagnetic ratio. The fact that the average of the red/green data points coincides with $\nu_\mathrm{H}$ and the blue data points with $2\nu_\mathrm{H}$ is a strong indication that the ESEEM is caused by dipolar coupling to protons. This is not surprising given the significant amount of water in the lattice of [HoW$_{10}$]$\cdot n$H$_2$O ($n\approx 35$ in fully solvated crystals). Indeed, a strong proton ESEEM effect is expected in this field range where the Ho-H dipolar coupling strength is comparable to the proton Larmor frequency (see below). By contrast, all other nuclei are predominantly non-magnetic, either due to low $\gamma$-values or low abundance of magnetic isotopes.

A qualitative understanding of the ESEEM spectrum is obtained by first considering the simplest possible case of coupled $S = \frac{1}{2}$ and $I = \frac{1}{2}$ spins in the high-field limit in which $\nu_\mathrm{H} \gg A$, where $A$ (=~$A_{zz}/h$, $A_{zz}$ is the $z$-component of the hyperfine tensor) quantifies the bare dipolar coupling  strength in frequency units. ESEEM arises due to excitation of formally forbidden zero- and double-quantum transitions that rotate coupled electron and nuclear spins.\cite{StollGoldfarbBook} The modulation results from combinations of the allowed ($\nu_a=\gamma_e B_0\pm\frac{1}{2}A$, $\gamma_e$ is the electron gyromagnetic ratio) and formally forbidden ($\nu_f=\gamma_e B_0\pm\nu_\mathrm{H}$) transition frequencies at: $|\nu^\pm_a - \nu^\mp_a| = A$, $|\nu^+_f - \nu^\pm_a| = |\nu^-_f - \nu^\pm_a| = \nu_\mathrm{H}\pm\frac{1}{2}A$ and $|\nu^\pm_f - \nu^\mp_f| = 2\nu_\mathrm{H}$.\cite{StollGoldfarbBook} One may then understand the red/green data points in Fig.~\ref{exp}(c) as being due to the hyperfine coupled proton frequencies, $\nu_\mathrm{H}\pm\frac{1}{2}A^{\mathrm{eff}}$, where $A^{\textrm{eff}}$ is an effective coupling strength on account of the new physics that emerges at the CT ($A^{\textrm{eff}}$ is further renormalized for HoW$_{10}$ due to the fact that $S \neq \frac{1}{2}$). Crucially, $A^{\mathrm{eff}} \rightarrow 0$ at the CT, which may be understood as being a consequence of the effective electron gyromagnetic ratio, $\gamma^{\mathrm{eff}}_e$, crossing through zero at $B_0 = B_{\textrm{min}}$ ($\gamma^{\mathrm{eff}}_e\propto df/dB_0$ or $\langle\hat{S}_z\rangle$, the $z$-component spin expectation value), as illustrated in Fig.~\ref{Zeeman}(b), where the ESR (clock) frequency couples quadratically to $B_0$ at the avoided crossing (CT), in contrast to the usual linear coupling far from the CT. This is why the ordering of red and green circles switches at the CT, i.e., there is a smooth evolution of $A^{\mathrm{eff}}$ ($\propto\gamma^{\mathrm{eff}}_e$) such that it switches sign at the CT. Remarkably, to first order, this implies that the effective dipolar coupling to protons vanishes right at the CT; hence the ESEEM effect also vanishes at the CT, as does the electron-nuclear decoherence, leading to the steep rise in $T_\mathrm{m}$ as one approaches the CT [$=8.43(6)~\mathrm{\mu}$s @ the CT].\cite{Shiddiq2016} Meanwhile, the ESEEM modulation depth grows with the detuning, $\Delta B$ (i.e., with $\gamma^{\mathrm{eff}}_e$), away from the CT, as does the electron-nuclear contribution to the central spin decoherence, i.e., $T_\mathrm{m}$ decreases to $\sim$$1~\mathrm{\mu}$s far from the CT.\cite{Shiddiq2016}

The ESEEM effect is ultimately governed by collective coupling of the Ho$^{\mathrm{III}}$ ion to the entire nuclear bath. However, the $1/r^3$ dependence of the dipolar interaction and large value of $\gamma_{\mathrm{H}}$ in comparison to other nuclei results in a spectrum that is dominated by nearby protons,\cite{Stoll20,Chen20} the closest of which are $\sim$4~${\mathrm{\AA}}$ from the central Ho$^{\mathrm{III}}$ ion.\cite{Liu2021} At this separation and in the linear Zeeman regime [$\Delta B_0 > 200$~mT from the CT in Fig.~\ref{Zeeman}(b)], the maximum Ho-H dipolar coupling strength, $A^\mathrm{max} \approx 3$~MHz (=~$2\mu_{\mathrm{o}}\mu_{\mathrm{Ho}}\mu_{\mathrm{H}}/4\pi h r^3$); this assumes $m_J=\pm4$ for the ground state of Ho$^{\mathrm{III}}$. The experimental results displayed in Fig.~\ref{exp} remain very far from this linear regime, which is why the separation of the red and green data points ($A^{\mathrm{eff}} \approx 0.4$~MHz) is well below the maximum. Meanwhile, ESEEM measurements far from the CT are hampered by the short phase memory time. Nevertheless, one would expect to observe an ESEEM effect in this field range because $A^{\mathrm{eff}}$ is of the same order as the proton Larmor frequency, $\nu_H = 1.1$~MHz at 25.5~mT. Indeed, ESEEM is also observed at the 2$^{\mathrm{nd}}$ ($\nu_H = 3.3$~MHz) and 3$^{\mathrm{rd}}$ ($\nu_H = 5.4$~MHz) CTs. Although the effect is less pronounced, the same qualitative behavior is found, i.e., a vanishing of ESEEM at each CT and harmonic content centered at $\nu_{\mathrm{H}}$ and $2\nu_{\mathrm{H}}$. Therefore, the enhanced coherence in the vicinity of the CTs provides a window through which to observe the ESEEM, which ultimately vanishes right at the CT because $\gamma^{\mathrm{eff}}_e \rightarrow 0$. We note that no modulation is discernible at the 4$^{\mathrm{th}}$ CT ($\nu_H = 7.6$~MHz), presumably because the effective dipolar coupling is just too weak in comparison to $\nu_\mathrm{H}$.

In order to gain microscopic understanding, we developed a simplified Hamiltonian for a central electron spin coupled to a finite proton spin bath. In order to preserve computational resources for the bath, we model the electronic system as an $S = 1$ spin with longitudinal and transverse anisotropy [Fig.~\ref{Zeeman}(a)]:
\begin{align}\label{ZH}
\hat{H}_S = D [\hat{S}_z^2 - \tfrac{1}{3} S (S + 1)] + E (\hat{S}_x^2 - \hat{S}_y^2) + \gamma_e(B_0-B_\textrm{min})\hat{S}_z\,,
\end{align}
where $\hat S_j$ are spin-1 generators of rotation about axis $j$, while $D$ and $E$ are the 2\textsuperscript{nd} order axial and rhombic zero-field splitting (anisotropy) parameters, respectively. $B_\textrm{min}$ is introduced to shift the CT away from $B_0 = 0$, mimicking the effect of the on-site hyperfine interaction with the $^{165}$Ho nuclear spin; note that this field does not act on the proton bath. The eigenvectors of Eq.~(\ref{ZH}) at the CT (i.e., when $\Delta B = 0$) are $|\pm\rangle=\tfrac{1}{\sqrt{2}}(|\uparrow\rangle\pm|\downarrow\rangle)$ and $|0\rangle$, with energies $-\tfrac{1}{3}|D| \pm E$ and $+\tfrac{2}{3}|D|$, respectively. Here, $|\uparrow\rangle$, $|\downarrow\rangle$, and $|0\rangle$ are the states with $\langle \hat S_z\rangle=\pm1$ and $\langle \hat S_z\rangle=0$, respectively. 

We set $D=-45$~GHz, $|E| = 4.5$~GHz and $B_\textrm{min} = 23.5$~mT in order to mimic the actual low-energy electronic structure of HoW$_{10}$. These parameters ensure the same CT frequency, $\Delta = 2E = 9$~GHz, the same curvature of the two lowest lying levels, and a sizeable separation to the $|0\rangle$ state (Fig.~\ref{Zeeman}). As an aside, because $|\pm\rangle$ are energetically well-separated from $|0\rangle$ in the vicinity of the CT, we can project onto the two-dimensional subspace defined by the former, wherein,
\begin{eqnarray}
&&\hat S_z^2\rightarrow\mathbb 1\,,\,\,\hat S_z\rightarrow\sigma_x\,,\,\,\hat S_x^2-\hat S_y^2\rightarrow\sigma_z\,,\,\,\{\hat S_x,\hat S_y\}\rightarrow2\sigma_y\,,\,\,\nonumber\\
&&
\hat S_x\rightarrow0\,,\,\, \hat S_y\rightarrow0\,,\,\, \{\hat S_y,\hat S_z\}\rightarrow0\,,\,\, \{\hat S_z,\hat S_x\}\rightarrow0\,.
\label{proj}
\end{eqnarray}
\noindent{Using this notation, the Hamiltonian reduces to $\hat H_S\rightarrow E\sigma_z + \gamma \Delta B \sigma_x$, which precisely maps onto a `fictitious' spin-$\frac{1}{2}$ model subjected to an effective magnetic field in the $xz$-plane.\cite{Vega77} The eigenvectors, which are quantized along the effective field direction, are still denoted $|\pm\rangle$, although these are no longer equally weighted mixtures of $|\uparrow\rangle$ and $|\downarrow\rangle$ upon detuning from the CT. Nevertheless, at the CT ($\Delta B = 0$), one may visualize qubit operations within this subspace in terms of pure rotations around the $j$\textsuperscript{th} axis of the Bloch sphere defined by $|\pm\rangle$, according to the Pauli matrices, $\sigma_j$; the corresponding spin-1 operators are then easily found from Eq.~(\ref{proj}). This mapping is helpful in understanding the simulated Hahn-echo sequence (see Methods), as there is no simple analogy to the $S=\frac{1}{2}$ rotating frame for the actual $S=1$ spin dynamics.} 

The nuclear spin bath, which ultimately causes decoherence and the observed ESEEM effect, is described by $N$ protons coupled via dipolar interactions to the central $S = 1$ state,
\begin{align}\label{7_dipole-dipole}
\hat{H}_{SI} = \hat{S}_z\sum_{m=1}^{N}\left[A_{sc}^m\hat{I}_z^m + A_{psc}^m(\hat{I}^m_x+\hat{I}^m_y)\right]\,.
\end{align}
Here, we employ secular ($sc$) and pseudosecular ($psc$) approximations with phenomenological couplings $A_{sc}^m$ and $A_{psc}^m$, respectively; the $\hat I_j^m$ are generators that rotate the spin of the $m$\textsuperscript{th} proton around axis $j$. The pseudosecular interaction is often ignored due to averaging brought about by the mismatch in the proton Larmor and hyperfine frequencies. However, as previously discussed, this is not the case at the first CT. Indeed, the pseudosecular interaction turns out to be essential to the ESEEM effect because it is responsible for driving formally forbidden nuclear transitions during the Hahn echo sequence.\cite{StollGoldfarbBook} Meanwhile, the protons also undergo their own dynamics, independent of the central spin, according to
\begin{align}\label{heisenberg}
\hat{H}_{I}= -\sum_{m\neq n}D_{mn}(3\cos^2\theta_{mn} - 1)[2\hat{I}_z^m\hat{I}_z^{n} - \hat{I}_x^m\hat{I}_x^{n} - \hat{I}_y^m\hat{I}_y^{n}]-\gamma_\mathrm{H}B_0\sum_{m=1}^{N} \hat{I}_z^m\,.
\end{align}
\noindent{That is, each proton in the bath undergoes Larmor precession at a bare frequency $\gamma_\mathrm{H}B_0$, and couples to other protons via a dipolar interaction of strength $D_{mn}$~($\sim$10~kHz); $\theta_{mn}$ is the angle between \boldmath${B}$\unboldmath$_0$ and the vector joining protons $m$ and $n$. Energy conserving proton flip-flop processes, driven by the $(\hat{I}_x^m\hat{I}_x^{n} + \hat{I}_y^m\hat{I}_y^{n})$ term, are central to the electron spin decoherence process.\cite{Wolfowicz2013,Prokofiev00,Stoll20,Chen20} To simulate the ESEEM, we numerically recreate the two-pulse Hahn echo sequence \emph{in silico} by performing a time evolution according to the total Hamiltonian, $\hat{H}_{tot} = \hat{H}_S + \hat{H}_{SI} + \hat{H}_I$ (see Methods).}

\begin{figure}
\centering
\includegraphics[width=0.7\linewidth,clip]{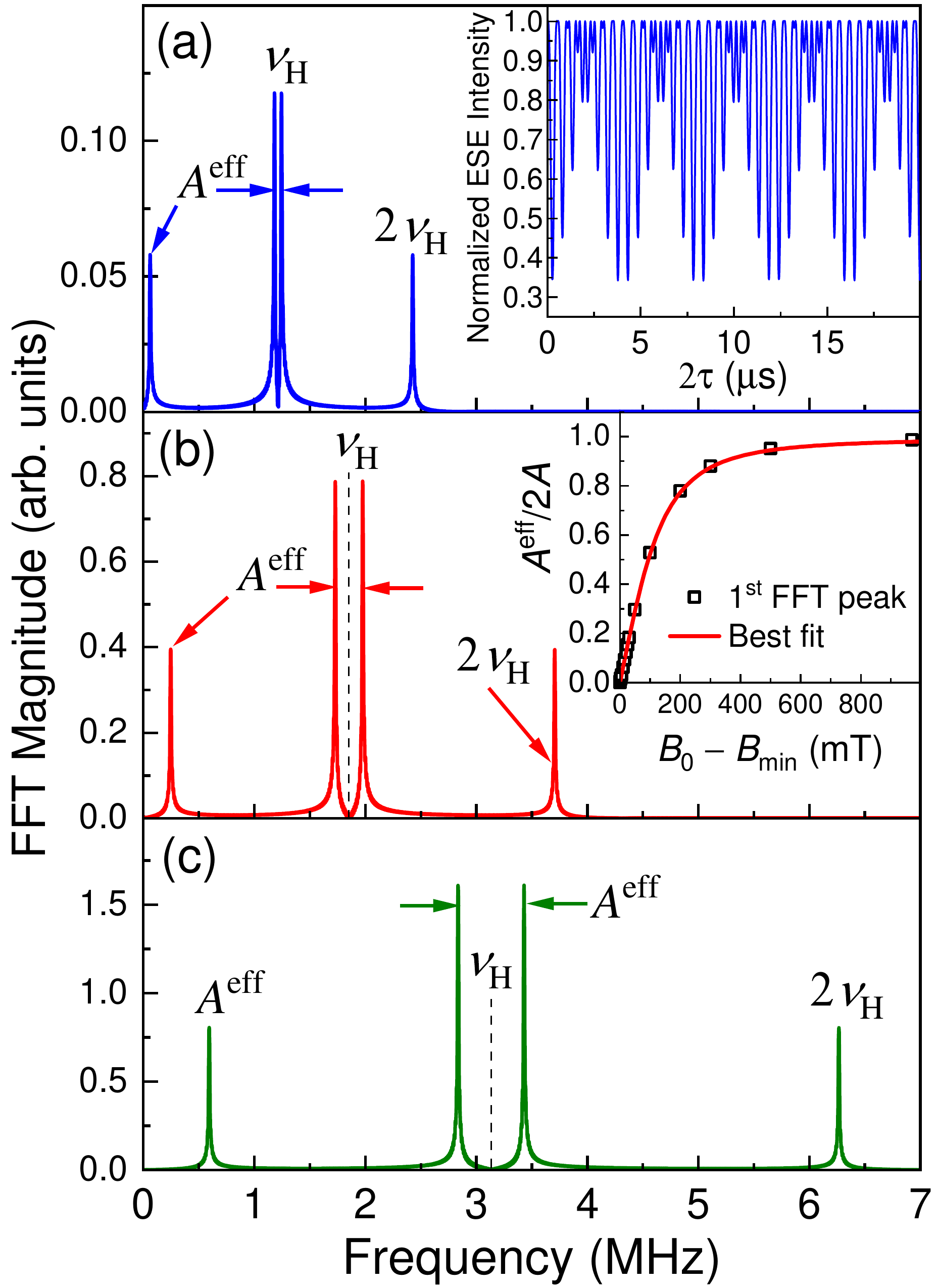}
\caption{FFTs of Hahn echo simulations for the simple case of a single proton ($N=1$) coupled to the central electron spin (see text for employed parameters) for different detunings, $B_0 - B_{\mathrm{min}}$ = +5~mT (a), +20~mT (b) and +50~mT (c); the inset to (b) shows a representative ESE intensity time trace. Several relevant frequencies are labeled in the FFT spectra. The inset to (a) plots $A^{\mathrm{eff}}$ deduced from the first FFT peak versus $B_0 - B_{\mathrm{min}}$; the red curve is a simple fit that assumes $A^{\mathrm{eff}} \propto df/dB_0$ from Fig.~\ref{Zeeman}(b).}\label{one_spin}
\end{figure}

As a warm up, we first consider the simple case of a single proton ($N=1$) coupled to the central $S=1$ spin, with $A=A_{sc}=2A_{psc} = 1$~MHz. Fig.~\ref{one_spin} displays FFTs of the Hahn echo simulations for several detuning fields [inset to (a) displays a representative time trace]. In analogy to the $S = \frac{1}{2}$ case, we associate the lowest frequency FFT peak, and the splitting of the peaks either side of $\nu_\textrm{H}$, with the effective hyperfine interaction strength, $A^{\mathrm{eff}}$; the inset to Fig.~\ref{one_spin}(b) plots this frequency as a function of $B_0 - B_{\textrm{min}}$. As can clearly be seen, and in analogy with the experiments, $A^{\mathrm{eff}} \rightarrow 0$ at the CT; indeed, the modulation (not shown) is also zero at the CT. Moreover, far from the CT, such that $\gamma_e |B_0-B_\textrm{min}|\gg |E|/h$, $A^{\mathrm{eff}} \rightarrow 2A$; the factor of two is due to renormalization because $S=1$ as opposed to $\frac{1}{2}$. Thus, in the high-field limit, FFT peaks occur at $2A$, $\nu_\mathrm{H}\pm A$ and $2\nu_\mathrm{H}$. Superimposed on the data in the inset to Fig.~\ref{one_spin}(b) is a phenomenological fit that assumes $A^{\mathrm{eff}} \propto \gamma_e^{\mathrm{eff}}$, deduced from $df/dB_0$ via Eq.~(\ref{ZH}). This confirms the idea that the variation in $\gamma_e^{\mathrm{eff}}$ (or $\langle\hat{S}_z\rangle$) in the vicinity the CT governs the dipolar coupling of the central spin to the nearby proton. The final thing to note from the inset to Fig.~\ref{one_spin}(a) is the absence of decoherence, i.e., the peak ESE intensity does not decay. This is because the two-spin system executes perfectly coherent coupled dynamics, with no quantum phase leakage, i.e., there is no bath associated with this model.

\begin{figure}
\includegraphics[width=1.0\linewidth,clip]{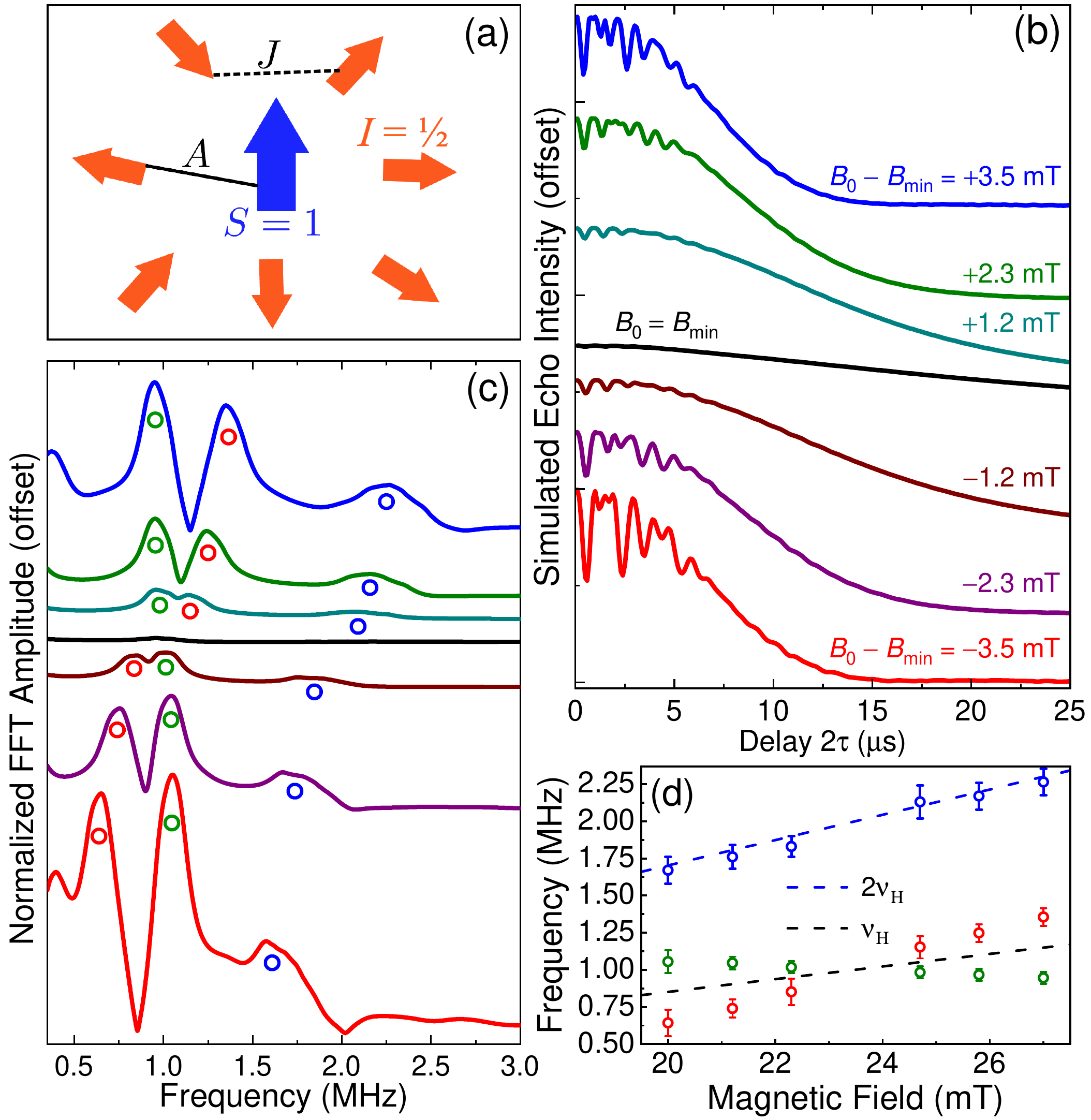}
\caption{(a) Schematic of a central spin coupled to $N=7$ nuclear spins, with $\langle A \rangle = 8$~MHz. (b) Simulated ESEEM time traces as a function of detuning, $B_0 - B_{\mathrm{min}}$. (c) FFTs of the time traces in (b). (d) Centers of the main FFT peaks in (c) as a function of $B_0$; the circles in (c) and (d) are colored according to the same scheme as Fig.~\ref{exp}.}\label{2in1}
\end{figure}

In order to better capture the physics associated with the spin bath, we extend the model to $N=7$ nuclear spins with a distribution of dipolar couplings to the central spin [(Fig.~\ref{2in1}(a)], enabling simulations of the ESEEM on reasonable timescales whilst also capturing the emergence of decoherence; we set $\langle{A^m_{sc}}\rangle = 2\langle{A^m_{psc}}\rangle = 8$~MHz to best reproduce the experimental results (see Methods for further details). Time traces for several detunings either side of the CT are displayed in Fig.~\ref{2in1}(b). As can be seen, the simulations qualitatively reproduce the experimental results in Fig.~\ref{exp}. A very clear ESEEM effect is observed that more-or-less vanishes at the CT. Moreover, the modulation depth increases with the detuning, $\Delta B$. The time traces also exhibit a very apparent decay in the coherence of the central spin dynamics, with a phase memory time, $T_\mathrm{m}$, that clearly diverges at the CT, i.e., the finite spin bath model causes decoherence of the central spin. Remarkably for such a simplified model, even the decoherence timescale is of the same order as the experiments. The only exception is at zero detuning, where the numerical decay is considerably flatter than the experiments. The residual decoherence observed at the CT in experiments is attributed to spin-lattice relaxation,\cite{Shiddiq2016} which is not included in our model; we comment on this further below. Fourier tranforms of the numerical time traces are displayed in Fig.~\ref{2in1}(c). Again, agreement with experiment is remarkably good. Indeed, a plot of the center frequencies of the main FFT peaks as a function of detuning, $B_0 - B_{\mathrm{min}}$ [Fig.~\ref{2in1}(d)], reveals identical behavior to the experiments, i.e., a pair of peaks at $\nu_\mathrm{H} \pm \frac{1}{2} A^{\mathrm{eff}}$ and a higher frequency peak at $\sim 2\nu_\mathrm{H}$; the peaks have been color coded in the same way as in Fig.~\ref{2in1}(c). Once again, it can be seen that $A^{\mathrm{eff}} \rightarrow 0$ at $B_0 = B_{\mathrm{min}}$, and increases with detuning from the CT.

The present experimental and theoretical investigation clearly demonstrates effective decoupling of an electron spin qubit from the surrounding nuclear bath at a CT, going beyond previous studies that simply show evidence for enhanced coherence.\cite{Shiddiq2016,Liu2021} In fact, our simulations reveal a pronounced enhancement in $T_{\mathrm{m}}$ at the CT, whereas the experiments on HoW$_{10}$ indicate that coherence is limited there by other factors. The primary culprit is spin-lattice ($T_1$) relaxation.\cite{Shiddiq2016} In particular, molecular vibrations that couple directly to the CF interactions(s) responsible for the CT (Fig.~\ref{Zeeman}) may be expected drive spin-lattice relaxation,\cite{Liu2021,Kragskow22,Ullah22} an effect not included in our model. However, weak decoherence is observed even at the CT in our numerical simulations. We attribute this to $2^{\rm{nd}}$-order coupling, $d^2 f/dB_0^2 = \gamma_e^2/\Delta$, i.e., $df/dB_0$ vanishes only precisely at the CT, and the HoW$_{10}$ qubit is therefore exposed to weak $^1$H dipolar field fluctuations either side of $B_{\mathrm{min}}$. This suggests that electron-nuclear decoupling should improve upon increasing the CT frequency, since the $2^{\rm{nd}}$-order coupling scales inversely with $\Delta$.

Electron spin-spin interactions have also been omitted from our model, since we consider only one Ho$^\mathrm{III}$ ion. One may expect the secular part of this interaction (i.e., $\hat{S}^m_z \hat{S}^n_z$) to decouple at a CT in exactly the same way that the proton bath decouples in this study, provided that the interaction is not too strong. As noted above, perfect decoupling occurs only to first-order ($df/dB_0 \rightarrow 0$) at the CT. However, $2^{\rm{nd}}$-order coupling should be weak if the spin-spin interaction strength is substantially weaker than the CT frequency ($\Delta=2E$),\cite{Wolfowicz2013,Kundu21} as is the case for well-separated ($>$nm) qubits. Meanwhile, although one may safely ignore angular momentum conserving electron-nuclear dipolar flip-flop processes in the present work because of the vastly different CT ($\Delta$) and proton Larmor ($\gamma_\mathrm{H}B_0$) frequencies, this is not the case for electron spin-spin interactions. Dipolar coupling within arrays of nominally identical qubits will cause decoherence due to flip-flop processes between resonant electron spins ($\Delta_1 = \Delta_2$) via the $\hat{S}^m_x \hat{S}^n_x + \hat{S}^m_y \hat{S}^n_y$ interaction.\cite{Wolfowicz2013,Escalera19} Correctly modeling this physics is more challenging, requiring a much larger bath with resonant and non-resonant qubits, due both to disorder (distributions in $\Delta$) and a dynamic distribution of dipolar interactions within the ensemble. Such a model contains complex many-body physics that lies outside of the realm of the present investigation.

One may anticipate that future QI devices based on molecular spins will feature controllable entangling interactions between individual qubits.\cite{Gaita-Arino2019} Crucially, this control would enable mitigation of resonant electron-electron spin flip-flop processes. Likewise, quantum sensing applications involving single qubits are immune to this mode of decoherence. However, it is virtually impossible to remove all sources of magnetic noise (particularly due to the nuclear bath), whilst maintaining the flexibility that molecular design principles allow. The present investigation therefore highlights the importance of CTs for suppressing electron-nuclear spin-spin decoherence. Moreover, one may expect these principles to apply quite generally to any type of CT. In this regard, hyperfine CTs show the most promise due to weaker coupling to molecular vibrations.\cite{Kundu21}

\section*{Methods}
\subsection{Experimental details.}
Since extensive discussions of sample preparation and handling, experimental setup and conditions, as well as the electronic properties that give rise to CTs in HoW$_{10}$ have been presented previously,\cite{Ghosh2012,Shiddiq2016} only brief descriptions of essential details are given here. Single crystals of Na$_9$[Ho$_{0.001}$Y$_{0.999}$(W$_5$O$_{18}$)$_2$]$\cdot n$H$_2$O were prepared according to the method described in Ref.~[\citen{AlDamen09}]. ESEEM measurements were performed using a commercial Bruker E680 X-band spectrometer equipped with a cylindrical TE$_{011}$ dielectric resonator (model ER 4118 X-MD5, with an unloaded center frequency of 9.75~GHz), which was overcoupled to increase bandwidth and, thus, allow measurements at frequencies down to 9.1~GHz.\cite{Shiddiq2016,Kundu21} The sample temperature was controlled using an Oxford Instruments CF935 helium flow cryostat and ITC503 temperature controller. 

All of the data presented in this study (Fig.~\ref{exp}) were obtained for a single crystal. However, similar ESEEM behavior has been observed in experiments performed on many other crystals of varying Ho$^{\rm{III}}$ concentration.\cite{Shiddiq2016} Although \emph{in situ} rotation of the crystal about a single-axis is possible, the low symmetry HoW$_{10}$ structure\cite{Ghosh2012} and the need for rapid sample loading to avoid degradation due to solvent loss resulted in an $\sim$22.5$^\circ$ misalignment between the applied field and the molecular Ising axis. This simply leads to a re-scaling of the CT fields: in this study, the lowest field CT occurs at $B_0 = B_{\mathrm{min}}=25.5$~mT, which is equivalent to a longitudinal field, $B_{0z}=23.6$~mT, where the $z$-direction defines the approximate HoW$_{10}$ four-fold symmetry axis. ESEEM results were derived from ESE decay curves generated using a standard two-pulse Hahn-echo sequence ($\pi/2 - \tau - \pi - \tau -$~echo) as a function of detuning from the CT field, $B_{\mathrm{min}}$. The frequency domain plots in Fig.~\ref{exp}(b) were obtained by performing FFTs of the time traces, which were zero padded by twice the number of data points and further smoothed using a 5 point average.

The spin Hamiltonian of the HoW$_{10}$ molecule may be described in terms of set of axial CF parameters, $B_k^q$ ($k=2,4,6$, representing the rank of the associated CF operator, $\hat{O}_k^q$, and $q=0$ the rotational order). Distortions away from the approximate $D_{4d}$ point symmetry of the HoW$_{10}$ molecule engage the tetragonal CF interaction, $B_4^4\hat{O}_4^4\propto (\hat{S}_+^4 + \hat{S}_-^4)$,\cite{Ghosh2012} which is effective in generating avoided crossings between the eight hyperfine sub-level pairs associated with the $m_J=\pm 4$ ground doublet, resulting in CTs at magnetic fields, $B_{\mathrm{min}} = \pm 23.6$, $\pm 70.9$, $\pm 118.1$ and $\pm 165.4$~mT (for an applied field, $B_0$, parallel to the molecular $z$-axis).\cite{Shiddiq2016} The W and O nuclei in the HoW$_{10}$ molecular core are predominantly non-magnetic, with the exception of $^{17}$O ($I=\frac{5}{2}$, $\gamma = 5.77$~MHz/T) and $^{183}$W ($I=\frac{1}{2}$, $\gamma = 1.77$~MHz/T) with $0.04\%$ and $14.3\%$ natural abundance, respectively. Moreover, their associated $\gamma$-values, along with those of the more distant $^{23}$Na and $^{89}$Y nuclei (both $100\%$ abundance) are considerably smaller than those of the proton. Consequently, one would not expect to see strong ESEEM from coupling to these other nuclei, i.e., assignment of the observed ESEEM to protons is unambiguous.

\subsection{Theoretical details.}
The two-pulse Hahn echo sequence was recreated \emph{in silico} by performing a time evolution according to the total Hamiltonian, $\label{one_nuclear_Ham} \hat{H}_{tot} = \hat{H}_S + \hat{H}_{SI} + \hat{H}_I$.\cite{Chen20} The initial density matrix at thermal equilibrium was defined in the lab frame as
\begin{align}\label{DM}
\rho_{eq} = \frac{\exp(-\beta \hat{H}_{tot})}{\mathrm{Tr}(\exp(-\beta \hat{H}_{tot}))},
\end{align}
\noindent{where} $\beta=h/k_B T$ and $T=5$~K. Instantaneous $\pi/2$ and $\pi$ pulses were performed according to the procedure described in the following paragraph. The density matrix was then allowed to evolve according to $\hat{H}_{tot}$ for a time interval $\tau$ after each pulse. Finally, the echo intensity was evaluated by computing the expectation value of the $z$-component of the Ho$^{\mathrm{III}}$ magnetization in the lab frame, $\textrm{Tr}_I (\rho \hat{S}_z )$, with the trace taken over the nuclear states. Exact matrix diagonalization demands considerable computational resources. Therefore, in order to carry out these calculations on reasonable time scales, a number of compromises were necessary. Foremost among these was the limitation on the size of the spin bath to $N=7$ protons. Meanwhile, based on \emph{a priori} knowledge of the spin dynamics, we could also optimize the time step and duration of the simulations, i.e., the time step (100~ns) results in a frequency cut-off, which we set to well above the $2\nu_{\mathrm{H}}$ frequency seen in the experimental spectra (Fig.~\ref{exp}), and the duration (100~$\mathrm{\mu s}$) was chosen to ensure a FFT resolution comparable to the experiments.

As discussed in the main text, the low energy $|\pm\rangle$ eigenvectors at the CT are not the usual $|\uparrow\rangle$ and $|\downarrow\rangle$ states relevant to the $S = \frac{1}{2}$ case; indeed, there is no simple rotating frame analogy that can easily be visualized in the case of the `real' $S=1$ system. One must therefore take care applying appropriate $\pi/2$- and $\pi$-pulses in order to generate the echo. In fact, one may reduce the problem to the simple Bloch sphere picture via projection onto the two-dimensional $|\pm\rangle$ subspace according to Eq.~(\ref{proj}), i.e., a `fictitious' spin-$\frac{1}{2}$ subjected to an effective magnetic field in the $xz$-plane ($\hat H_S \rightarrow E\sigma_z + \gamma \Delta B \sigma_x$).\cite{Vega77} The appropriate pulses can then be implemented via rotations about any axis that is perpendicular to the effective field, $\Vec{B}^{eff}$ ($=\sqrt{E^2 + (\gamma \Delta B)^2}$). Exactly at the CT ($\Delta B = 0$), where $\Vec{B}^{eff} \parallel z$, this is easily achieved using either the pure $\sigma_x$ or $\sigma_y$ Pauli matrix, corresponding to the spin-1 operators $\hat S_z$ and $\{\hat S_x,\hat S_y\}$, respectively. Away from the CT, $\Vec{B}^{eff}$ tilts towards $x$ within the $|\pm\rangle$ subspace. We therefore employ a pure $\sigma_y$ rotation, which does not depend on the orientation of $\Vec{B}^{eff}$, i.e., we implement pulses of the form $\exp[i\phi\{\hat S_x,\hat S_y\}/2]$, where $\phi$ denotes the rotation angle in radians. Although the $\{\hat S_x,\hat S_y\}$ operator has no direct correspondence with the microwave $B_1$ field employed in the experiments, it conveniently achieves the desired result. Moreover, it is formally equivalent to operating with $\hat S_z$ at the CT, which does correspond directly to the experimental parallel mode $B_1$ field. However, upon moving away from the CT, the ideal magnetic $\hat S_z$-pulse evolves with the applied field, $B_0$, as the eigenvectors acquire unequal $|\uparrow\rangle$ and $|\downarrow\rangle$ weights. Indeed, the durations of the $\pi/2$ and $\pi$ pulses employed in the real experiments had to be optimized for each field step, something that could be avoided in the simulations by implementing pure $\sigma_y$ rotations.

Additional subtleties of the calculations concern the precise details of the microscopic interactions. For example, in order to recreate a realistic proton bath, a distribution of electron-nuclear hyperfine coupling strengths, $A^m=A_{sc}^m=2A_{psc}^m$ ($m=1$~to~$N$), was generated with random values in the range from 7 to 9~MHz such that $\langle A \rangle =8$~MHz. Likewise, a distribution of proton-proton dipolar interactions was implemented by fixing the coupling strength in Eq.~(\ref{heisenberg}), $D_{mn} = \mu_{\mathrm{o}} \mu_{\mathrm{H}}^2 /8 \pi h r^3 \approx 10$~kHz ($\equiv 1.8$~$\mathrm{\AA}$ distance), and randomizing the angle $\theta_{mn}$. To compensate for the small size of the nuclear bath, the simulations were repeated ten times for different $A^m$ and $\theta_{mn}$ randomizations, then averaged; this approach is obviously vastly more efficient computationally compared to increasing the bath size. Not only do these measures better mimic the real [HoW$_{10}$]$\cdot n$H$_2$O system, they avoid the highly unphysical situation in which the seven protons are indistinguishable, with identical couplings to the central spin and to each other. The 8~MHz value for $\langle A \rangle$ was chosen so as to reproduce the effective hyperfine interaction seen in the experiments, i.e., splitting of the red/blue circles in Figs.~\ref{exp}c and \ref{2in1}d. This corresponds to a Ho-$^1$H separation of $\sim 2.9$~$\mathrm{\AA}$, which is a little under the known closest distance ($\sim 4$~$\mathrm{\AA}$), which we attribute to the fact that the model under-counts the number of nearby protons by about an order of magnitude, i.e., $n=35$ H$_2$O molecules, or 70 protons per Ho$^\mathrm{III}$ ion. Consequently, the smaller distance employed in the simulations effectively renormalizes the collective hyperfine coupling strength. Finally, the frequency-domain plots in Fig.~\ref{2in1}(c,d) were obtained by first subtracting a stretched exponential background from the time traces [Fig.~\ref{2in1}(b)], then performing FFTs of the residual ESEEM modulations; Fourier-transform filtering was used to smooth the resulting frequency-domain plots.

In spite of the aforementioned simplifying assumptions, the employed model captures the essential physics associated with the ESEEM effect in the vicinity of a realistic CT. Moreover, the simulations reproduce the experimentally observed electron-nuclear decoherence. Approximate cluster correlation expansion (CCE) methods are able to consider a much larger and more realistic bath consisting of thousands of protons. Indeed, such studies applied to simple spin-$\frac{1}{2}$ qubits (with no CT) obtain essentially perfect quantitative agreement with experimental phase memory times.\cite{Stoll20} However, they also reveal that decoherence is dominated by stochastic flip-flop processes associated with proton pairs that are relatively close (a few $\mathrm{\AA}$) to the central electron spin. As such, the exact quantum calculations considered here contain the same ingredients. It is therefore unsurprising that the obtained phase memory times agree with experiment to within approximately a factor of two. Indeed, the $N=7$ model enables exploration of many other microscopic aspects of the bath that influence decoherence. We hope to explore this further in the future. We wish to emphasize, however, that it was not our original intent to quantitatively reproduce the decoherence, but rather to qualitatively reproduce the ESEEM effect in the vicinity of a CT, something that this investigation has most definitely achieved.

\section*{Code availability}
The computer code to run our simulation is available from K. K. upon request.

\section*{Acknowledgements}
We are grateful to Eugenio Coronado and Shimon Vega for insightful discussion, and we thank Haechan Park and James Fry for assistance in estimating the closest Ho-$^1$H distances. The spectroscopic and theoretical work reported in this paper was supported by the Center for Molecular Magnetic Quantum Materials (M$^2$QM), an Energy Frontier Research Center funded by the US Department of Energy, Office of Science, Basic Energy Sciences under Award DE-SC0019330. Experimental work performed at the National High Magnetic Field Laboratory is supported in part by the National Science Foundation (under DMR-1644779) and the State of Florida. Synthesis of the HoW$_{10}$ sample was supported by: the EU (ERC-2018-AdG-788222 MOL-2D, the QUANTERA project SUMO, and FET-OPEN grant 862893 FATMOLS); the Spanish MCIU (grant CTQ2017-89993 and PGC2018-099568-B-I00 co-financed by FEDER, grant MAT2017-89528; the Unit of excellence ‘Maríade Maeztu’ CEX2019-000919-M); and the Generalitat Valenciana (Prometeo Program of Excellence).
 
\section*{Author information} 
\subsection{Contributions}
K.K. and J.C. performed the simulations. S. Hoffman developed the theory. The manuscript was written by J.C., J.M. and S. Hill with contributions from S. Hoffman, K.K., A.G. and X.-G.Z. The experiments were conceived and designed by S. Hill and D. K., while D.K. performed the measurements. Y.D. prepared the samples. J.S., A.G., X.-G.Z., S. Hill and H.-P.C. supervised the research.

\subsection{Corresponding authors}
Correspondence and requests for materials should be addressed to S. H., S. H., and H.-P. C. (email: silas.hoffman@ufl.edu, shill@magnet.fsu.edu, and hping@ufl.edu).

\section*{Ethics declaration} The authors declare that they have no competing financial interests.

\pagebreak

\end{document}